\documentstyle[preprint,aps,psfig]{revtex}
\begin{document}
\draft
\title{Turing patterns and solitary structures under global control}
\author{L.M. Pismen}
\address{ Department of Chemical Engineering,\\
Technion -- Israel Institute of Technology, 32000 Haifa, Israel} 
\date{}
\maketitle

\begin{abstract}
Striped Turing patterns and solitary band and disk structures are 
constructed using a three-variable multiscale model with cubic 
nonlinearity and global control. The existence and stability conditions 
of regular structures are analysed using the equation of motion of 
curved boundaries between alternative states of the short-range 
component. The combined picture of transitions between striped and 
spotted patterns with changing level of global control is in qualitative 
agreement with the results of the computer experiment by Middya 
and Luss \cite{luss93}.
\end{abstract}

\pacs{82.20.Mj, 05.70.Ln}

\section{Introduction}

Stationary patterns of chemical activity (Turing patterns), have 
attracted wide attention both as a key to understanding complex and 
unusual phenomena in chemical and electrochemical kinetics and as 
a possible basis of morphogenesis and biological diversification 
\cite{kur,field,mur}.  Similar patterns, emerging as a result of 
spontaneous symmetry breaking have been known and intensively 
studied long before that in fluid mechanics \cite{chan,crho}.

Reproducing stationary chemical patterns in the laboratory under 
controlled conditions turned out to be a rather elusive task, as most 
observations, such as the BZ reaction, produced propagating chemical 
waves rather than stationary structures; immobilization of chemical 
waves was achieved only due convective effects \cite{mul} or 
imposed spatial gradients in the Couette reactor \cite{swin,arn}.  In 
recent years, novel experimental techniques were applied to obtain a 
variety of complex chemical patterns in gel reactors 
\cite{cast,quy,quy2,lee} and on catalytic surfaces \cite{ertl}. A 
parallel progress in computing and visualization is evident in 2$d$ 
patterns obtained in recent computer experiments \cite{pea,luss93}. 

Most common models of large-amplitude dynamics, such as the 
FitzHugh -- Nagumo equation, favor propagating waves, rather than 
stationary patterns.  A basic mechanism generating stationary 
inhomogeneities is a combination of a short-range activator and a 
long-range inhibitor with a sufficiently fast dynamics \cite{seg}.  
Formation of chemical patterns can be understood analytically when 
characteristic spatial and temporal scales associated with different 
reactants and other dynamic variables, like temperature or potential, 
are widely separated; this is typical to heterogeneous reacting 
systems, due to great disparities in diffusion rates.  This scale 
separation allows to use rational approximation techniques, first of 
all, the method of matched asymptotic expansions, to build up and 
investigate dynamically complex non-equilibrium patterns 
\cite{fife,pi79}.

The basic element of a non-equilibrium pattern in a multiscale 
system is a propagating front, or kink, separating regions of 
prevailing alternative stationary states of the short-range subsystem, 
usually characterized by a low or high level of the activator. More 
complex patterns are constructed by allowing other constituents, 
varying on a longer spatial scale, to modify the motion of kinks and 
to mediate their interaction, thereby leading to the formation of 
bound pairs, trains, etc. Generically, the kinks are ``{\em fast}'', i.e. 
propagate with a speed of the same order of magnitude as a ``unit'' 
speed that can be conjectured from the characteristic length and time 
scales of the short-range activator. Wave patterns (with relaxation 
oscillations at any fixed point) are constructed as trains of fast kinks 
separated by distances far exceeding their width. 

Stationary or slowly evolving structures can be ``assembled'' in a 
similar way but using as basic elements ``{\em slow}'' kinks 
propagating with a speed much less than unity. The slowing-down 
can be either a consequence of a symmetry that is only weakly 
broken due to the influence of long-range fields, or a result of a 
dynamic adjustment to changing levels of a long-range variable. A 
Turing pattern can take the form of alternating domains separated 
by immobilized kinks.

Dynamics of kinks is sensitive to the influence of {\em global} 
variables which are present in some of the most common laboratory 
set-ups. In experiments with catalytic wires \cite{sch,luss91}, the 
global control is due to heating input regulated to keep the average 
wire resistance (i.e. essentially the average temperature) constant. If 
local kinetics is bistable, the global control of this kind causes 
spontaneous symmetry breaking on a global scale \cite{pi78} due to 
immobilization of kinks. In electrochemical experiments 
\cite{lev,lev2} or glow discharge devices \cite{purw}, the global 
controlling factor is the conservation of the total current. In both 
catalytic and electrochemical studies, oscillatory local kinetics, that 
would normally lead to a pattern of {\em propagating} waves (or 
coupled fast kinks) in a distributed system, generates {\em standing} 
waves under conditions of global control. Still more complex 
dynamics of pulses and kinks was obtained in numerical experiments 
\cite{luss92}.

A different kind of global control is connected with the influence of a 
rapidly mixed gas phase in catalytic reactions.  Recent model 
computations \cite{zou} indicated that a global gas-phase oscillation 
was the factor causing transition to standing waves in the 
experiments of the group of Ertl \cite{ertl}. Global interaction 
through the gas phase is usually detrimental to pattern formation, as 
it tends to synchronize the reaction on the entire surface. Recent 
computations by Middya and Luss \cite{luss93} demonstrated a 
sequence of transitions from striped to spotted patterns and, 
eventually, to restoration of spatial homogeneity with an increased 
level of global control. 

The main purpose of this communication is to find analytic criteria 
for such transitions in a reaction-diffusion system with separated 
scales. Our description of Turing patterns is mainly based on 
representing them as collections of ``solitary'' band or disk 
structures, although we shall also consider regular striped patterns. 
Different objects bounded by kinks can be indeed viewed as solitary 
even at moderate distances, since interaction of kinks decays 
exponentially with separation. 

The motion of a slow kink is influenced by its geometric form even 
when its curvature is small. This distinguishes them from fast kinks, 
which are insensitive to the geometry unless their curvature radius 
becomes comparable to their width. Solitary structures bounded by 
slow kinks are stabilized by a delicate balance between the 
geometric factors and the influence of long-range variables, and their 
existence and stability conditions are strongly dependent on their 
geometric shape.

In the following, we shall consider only slowly evolving structures 
that are built up of slow kinks, and have a characteristic scale far 
exceeding the width of the kink. The basic three-variable model with 
cubic nonlinearity is formulated in Section \ref{SBASIC}, and the 
equation of motion of a bent kink is derived in Section \ref{SINNER}. 
We proceed with the construction of various stationary structures in 
Section \ref{SREG}, and the analysis of their stability in Section 
\ref{SSTAB}. The combined picture of the existence and stability 
regions of solitary band and disk solutions, as discussed in the last 
Section, is in qualitative agreement with the results of the computer 
experiment by Middya and Luss \cite{luss93}.

\section{Basic equations\label{SBASIC}}

\subsection{Reaction-diffusion model with separated 
scales\label{rdmodel}}

\begin{figure}
\psfig{figure=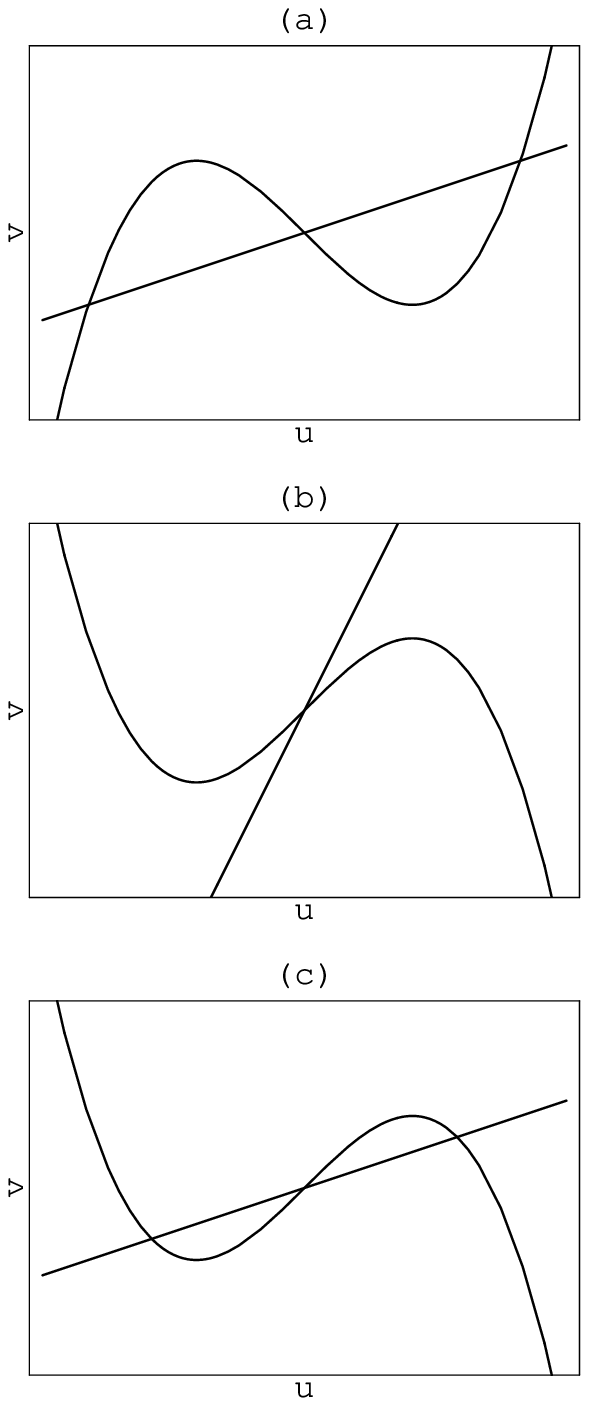}
\caption{Synclinal (a) and anticlinal (b,c) disposition of null isoclines.}
 \label{fuv}
\end{figure}

The influence of global control variables on the formation of 
stationary non-equilibrium patterns can be understood qualitatively 
within the common context of pattern formation in reaction-diffusion 
systems with separated scales. The common recipe for a stationary 
pattern \cite{seg,fife,pi79} calls for a pair of equations for two 
reacting and diffusing species: a short-range ``activator'' $u$ and a 
long-range ``inhibitor'' $v$:
     \begin{eqnarray}
 \gamma u_t & = & \epsilon^2 \nabla^2 u + f(u,v),  
      \label{ueq}   \\
   v_t & = & \nabla^2 v + g(u,v) .
      \label{veq}     \end{eqnarray}
The equations are written using the characteristic time and length 
scales of the long-range component as basic units. The small 
parameter $\epsilon$ is the square root of the ratio of diffusivities of 
the two species. It is supposed that the nonlinear function $f(u;v)$, 
with $v$ considered as a parameter, has three zeroes within a certain 
range of $v$, two of them, $u=u^\pm(v)$, being stable, and one 
unstable. Under these conditions, Eq.(\ref{ueq}) has propagating kink 
solutions that separate regions with prevailing ``lower'' and ``upper'' 
states. Everywhere except in an $O(\epsilon)$ vicinity of the kink, 
the dynamics of $v$ is defined by a closed reaction-diffusion 
equation
\begin{equation}
 v_t  = \nabla^2 v + h^\pm(v)
\label{oveq} \end{equation}
with a discontinuous source function given by alternative forms 
$h^\pm(v) = g(u^\pm(v), v)$ in alternative regions separated by a 
kink. It is clear that most of the information carried by the function 
$g(u,v)$ is irrelevant. Its only essential property is reflected by the 
disposition of null isoclines, i.e loci of zeroes of the functions $f(u,v)$ 
and $g(u,v)$ in the plane $(u,v)$, which can be either {\em 
synclinal}, as in Fig.~\ref{fuv}a, or {\em anticlinal}, as in 
Figs.~\ref{fuv}b,c. 

When the disposition is anticlinal, stable global stationary states 
satisfying $h^\pm(v)=0$ may either disappear altogether, as in 
Fig.~\ref{fuv}b, or become {\em excitable}, as in Fig.~\ref{fuv}c. An 
excitable state $u= u^\pm (v)$ retreats when placed in contact 
with an alternative steady state of the the short-scale equation $u= 
u^\mp (v)$ at the same value of $v$. An advancing front of the 
alternative state is arrested when the level of the inhibitor adjusts to 
shifts in the front position, and relaxes to the value $v=v_s$  
satisfying the stationarity condition for the kink:
\begin{equation}
I(v_s) \equiv \int_{u_-(v_s)}^{u_+(v_s)} f(u,v_s) du =0.
\label{integral}  \end{equation}
An immobilized kink-antikink pair forms a solitary structure on the 
background of an excitable state. Such structures were constructed 
with the help of a multiscale model with piecewise-linear kinetics 
\cite{ohta}, and, recently, detected experimentally in gel reactors 
\cite{lee}. A regular array of alternating kinks and antikinks forms a 
Turing structure, which is the only stable stationary state in the 
``oscillatory'' case corresponding to the disposition of null isoclines in 
Fig.~\ref{fuv}b. 

Stability of immobilized kink structures under conditions when the 
level of the inhibitor adjusts momentarily to shifts in the position of 
kinks can be supported by the following qualitative argument. We 
take note that, when the disposition of  null isoclines is anticlinal, as 
in Fig.~\ref{fuv}b,c, $v$ is depleted in the lower state $u^-$ and 
produced in the upper state $u^+$, while, on the contrary, the lower 
state advances at higher, and the upper, at lower values of $v$. 
Consider, for example, a finite region with $u=u^-$ immersed in a 
continuum with $u=u^+$. If it starts to shrink, the level of $v$ 
would raise and, as a result, the kinks bounding the region would be 
immobilized again (a more precise analysis, taking also into account 
the local gradient of $v$ and effects of the kink curvature is given in 
Section~\ref{SSTAB}). The same argument proves that immobilized 
fronts should be unstable when the disposition of null isoclines is 
synclinal, as in Fig.~\ref{fuv}a, so that the lower state advances at 
lower, and the upper, at higher values of $v$. 

\subsection{Global control\label{gmodel}}

The above qualitative arguments can be extended to the case when 
$v$ is a {\em global} controlling variable. The model with global 
interactions can be viewed as a limiting case of the model with 
separated scales, which is obtained when the diffusivity of the 
long-range variable tends to infinity, so that its diffusional length far 
exceeds the size of the system. Then Eq.~(\ref{veq}) is replaced by 
the integral equation
\begin{equation}
   v_t = \langle g(u,v) \rangle ,
\label{gleq}  \end{equation}
where $\langle \ldots \rangle$ denotes averaging over the entire 
reactive surface. The respective form of eq.~(\ref{oveq}) is
\begin{equation}
 v_t  = \alpha h^-(v) + (1-\alpha) h^+(v), 
\label{gloveq} \end{equation}
where $\alpha$ is the fraction of the surface occupied by the lower 
state. 

The action of either a long-range or a global inhibitor leads to the 
formation of stable inhomogeneous structures in the case of an 
anticlinal (but not a synclinal) disposition of null isoclines. The only 
distinction lies in the characteristic size of the inhomogeneities 
(either a wavelength of a regular pattern or a width of an excited 
domain). This size is, generally, determined by the diffusional length 
of the inhibitor, and increases to become comparable with the size of 
the system in the case of a global control. Under conditions of global 
control, the size of inhomogeneities is not limited, and the system 
evolves to minimize the length of the boundary between the domains 
with prevailing alternative states. Each disjoint domain tends to 
become circular as its circumference shrinks due to the line tension 
of a bounding kink. Further decrease of the length of the boundaries 
is caused either by the coalescence of domains, or by dissolution of 
smaller and growth of larger domains under the constraint of a 
conserved total area. This process is similar to spinodal 
decomposition in equilibrium systems described by the Cahn -- 
Hilliard equation \cite{cahn}. Its driving force is either the attractive 
interaction between like domains, which decays exponentially at a 
distance comparable with the diffusional length of the short-range 
variable, or the line tension, which causes the kinks to move with a 
velocity proportional to the ratio of the same diffusional length to the 
local curvature radius. As in equilibrium systems, the ``ripening'' 
process becomes exceedingly slow as the prevailing size of the extant 
domains increases. 

Separation into domains with prevailing alternative steady states in 
an anticlinal system is the basis of the ``global regulator'' model of 
spontaneous symmetry breaking \cite{pi78}. This mechanism works, 
for example, in a catalytic wire heated by electric current when the 
controls are adjusted to keep the average temperature within the 
unstable region between the lower (extinct) and higher (ignited) 
states. In this case, a simple suitable form of the model equations is
\begin{equation}
\gamma u_t  =  \epsilon^2 \nabla^2 u + F(u) + U , \;\;\;
   U = B (u_s -\langle u \rangle ),
\label{heateq}   \end{equation}
which corresponds to an anticlinal disposition. The short-range 
variable $u$ can be identified as temperature and the global variable 
$U$, as the overall heating rate. The disposition would 
not change qualitatively in a more precise model taking into account 
the dependence of local heat dissipation on local temperature.
 
The situation is different when the global control is caused by the 
interaction with a rapidly mixed ambient fluid phase that changes its 
composition and/or temperature in response to the changes of the 
averaged state of the catalytic surface. The simplest model suitable 
to this case is 
\begin{equation}
\gamma u_t  =  \epsilon^2 \nabla^2 u + F(u) + U , \;\;\;
   U_0-U = B (U -\langle u \rangle ),
\label{gaseq}    \end{equation}
where $U$ has now the meaning of concentration or temperature in 
the ambient fluid. In this case, the disposition of null isoclines is 
{\em synclinal}, and no symmetry breaking occurs. If other factors 
induce formation of Turing patterns or excitation domains on the 
catalytic surface, the global interaction of this kind acts to modify the 
emerging patterns, and may suppress them altogether when it is 
sufficiently strong. 

\subsection{Three-variable model with a cubic nonlinearity 
\label{cmodel}}

In the following, we shall consider a minimal model including three 
variables: a short-range activator $u$, a long-range inhibitor $v$, and 
a global controller of the synclinal type $U$ that suppresses pattern 
formation. Analytical studies of non-equilibrium patterns often 
employ equations with a discontinuous piecewise linear source 
function for a {\em short-range} variable, as this form of the source 
function makes computations most simple. Such a model was used, in 
particular, to construct stationary structures as immobilized pairs or 
arrays of kinks \cite{ohta}. In any physical situation discontinuous 
source functions can only appear when there is some hidden 
short-range variable that has been removed using a procedure of the 
same kind that has lead us from Eq.(\ref{veq}) to Eq.(\ref{oveq}). In 
the long-range equation, however, the position of the discontinuity is 
not free but has a dynamics of its own, dependent on the dynamics 
of the short-range variable in the kink region. If this dynamics is 
ignored, and a discontinuous function is used in the basic short-scale 
equation, it can lead to undesired consequences, like 
the divergency of the propagation speed near the discontinuity. 

The simplest physically admissible source function in the 
equation of $u$ should contain a cubic nonlinearity; the dependence 
on both $v$ and $U$ can be linear. We shall concentrate upon the 
case when the symmetry breaking between the alternative states of 
the short-range variable is weak, and causes the kinks to propagate 
with a speed comparable to that induced by a weak curvature with a 
radius far exceeding the diffusional length of the activator. This will 
allow us to take into account geometric factors while remaining 
within applicability limits of the multiscale perturbation scheme. 
The simplest suitable form of the model is
   \begin{eqnarray}
\gamma u_t & = & \epsilon^2 \nabla^2 u + (1 - u^2)u + \epsilon (U-v), 
      \label{sueq}   \\
   v_t & = & \nabla^2 v - v - \nu  + \mu u.
    \label{sveq}  \\
   U & = & U_0 + \frac{\beta}{\epsilon}( \langle u \rangle -1).
\label{sgleq}   \end{eqnarray}
With $U \le O(1)$, eq.(\ref{sueq}) is just the Fisher -- Kolmogorov 
equation supplemented by an $O(\epsilon)$ symmetry-breaking 
term. The latter is of the right order of magnitude to balance kinetic 
and geometric effects in the kink propagation. Both parameters 
$\mu$ and $\beta$ are supposed to be positive, which corresponds 
to the anticlinal and synclinal disposition of null isoclines, 
respectively, for $v$ and $U$. Under these conditions, spontaneous 
pattern formation triggered by $v$ is suppressed by the global 
variable which imitates the action of the environment in a mixed 
vessel. In the absence of global control ($\beta=0$), both steady 
states $u_\pm \approx \pm 1$ are weakly excitable at 
$|\nu|<\mu$, while at $|\nu|>\mu$ only one state is excitable. 

The dynamics of the global variable is supposed to be very fast, and 
to insure stable relaxation to the solution of Eq.~(\ref{sgleq}). The 
global balance condition (\ref{sgleq}) is written in such a way that, 
provided $\beta=O(1)$, the value of $U$ is large, and relaxes to an 
$O(1)$ level only when the upper state $u_+ \approx 1$ is prevailing, 
so that the fraction of the surface occupied by the lower state is at 
most of $O(\epsilon)$. Therefore stable patterns generated by this 
model at moderate values of $\beta$ can only take the form of a
solitary band or disk with $u<0$ immersed in a continuum with 
$u>0$. The only globally stable stationary state is, to the leading 
order, $u=1, \; v=\mu-\nu, \; U = U_0$. At $\beta <\frac{1}{4}$, 
another pair of solutions exists, one of which is unstable and the 
other one is metastable. 

\section{Equation of motion for a curved kink\label{SINNER}}

Problems with separated scales are generally solved using matched 
asymptotic expansions in the {\em inner} region (in the vicinity of 
the kink) and in the {\em outer} regions where one of the alternative 
equilibria of the short-range variable is approached. We shall derive 
the equation of motion for a slowly propagating and moderately bent 
kink as a solvability condition that insures the existence of a 
stationary solution of Eq.(\ref{ueq}) in an aligned frame comoving 
with the kink. 

The coordinate frame aligned with a planar curve is defined in the 
following way. Let the curve be parametrized by a coordinate $\eta$, 
and its position in the plane defined as ${\bf x}(\eta)$. 
The derivatives of this function define the unit tangent vector ${\bf 
t} = {\bf x}'(\eta)/|{\bf x}'(\eta)|$ and the curvature vector 
$d{\bf t}/d\eta = \kappa {\bf n}$, where ${\bf n}$ is the normal to 
the curve, and $\kappa$ is the scalar curvature. By convention, the 
$\xi$ axis has its origin on the curve and is directed oppositely to 
${\bf n}$. The coordinate lines $\xi=$ const are obtained by shifting 
the curve along the normal by a constant increment; evidently, this 
shift causes the length to increase on convex, and to decrease on 
concave sections of the curve. The length element is defined as $ds^2 
=d\xi^2 + (1+\kappa \xi)^2 d\eta^2$. Accordingly, the Laplacian is 
expressed as 
\begin{equation}
\nabla^2 = (1+\kappa \xi)^{-1} \left[ 
  \partial_\xi (1+\kappa \xi) \partial_\xi +
  \partial_\eta (1+\kappa \xi)^{-1} \partial_\eta \right] .
\label{lap} \end{equation}

The aligned system is well defined only sufficiently close to the 
curve, due to a singularity developing on the concave side at a 
distance of $O(\kappa^{-1})$. If the curvature radius far exceeds the 
the diffusional range of the short-scale variable $u$, the aligned 
system remains regular in the region where $u$ changes between 
the alternative equilibria $u=u_\pm$. In the vicinity of a kink, the 
coordinate $\xi$ has to be rescaled by the factor $\epsilon$ to 
accommodate the rapid change of $u$. Rescaling the tangential 
coordinate is not necessary as long as the curvature radius of the 
kink is of $O(1)$ on the extended scale, and far exceeds the kink 
width. Adopting this scaling, we rewrite the Laplacian Eq.(\ref{lap}) 
as
\begin{equation}
\nabla^2 = \epsilon^{-2} \partial^2_\xi + 
    \epsilon^{-1} \kappa \partial_\xi +
   \partial^2_\eta - \kappa^2 \xi \partial_\xi  +  O(\epsilon).
\label{lap1} \end{equation}

The time derivative is transformed in the comoving frame as 
$\partial_t \to \partial_t - C \epsilon^{-1} \partial_\xi$. The 
additional term matches the curvature term by the order of 
magnitude, provided the propagation velocity $C$ is scaled as 
$C=\epsilon^2 \gamma^{-1} c$ where $c=O(1)$. In the inner region, 
the long-range variable is constant across the kink in the leading 
order. Assuming a linear dependence on some combination $w$ of 
the long-range and global variables, as e.g. in Eq.~(\ref{sueq}), we set 
in Eq.~(\ref{ueq}) $f(u,v)= f_0(u) + \epsilon w f_1(u)$ where $f_0(u)$ 
is odd. Expanding the short-range variable as $u=u_0(\xi) + \epsilon 
u^{(1)}(\xi,\eta) + \ldots$ yields, in the zero order
\begin{equation}
 u_0''(\xi) + f_0(u_0) =0.
\label{u0eq} \end{equation}
This equation is verified by a stationary kink (antikink) solution 
satisfying the asymptotic condition $u_0(\xi) \to u^\pm$ at $\xi \to 
\pm\infty$ (for a kink) or at $\xi \to \mp\infty$ (for an antikink). 
In particular, for the cubic nonlinearity in Eq.(\ref{sueq}), the kink 
solution is
 \begin{equation}
 u_0(\xi) =  \tanh (\xi/\sqrt{2}).
  \label{kink} \end{equation}

The first-order equation is
\begin{equation}
u^{(1)}_{\xi\xi} + f_0'(u_0) u^{(1)} + \Psi(\xi,\,\eta) =0,  
\label{u1eq} \end{equation}
which contains the inhomogeneity
\begin{equation}
\Psi(\xi,\,\eta) = (c + \kappa) u_{0}'(\xi) + f_1(u_0) w.
\label{inhom}  \end{equation}
The propagation speed is determined by the solvability condition of 
Eq.(\ref{u1eq}), which implies the orthogonality of the Goldstone 
eigenmode $u_{0}'(\xi)$ of the homogeneous part, corresponding to 
the translational symmetry of a stationary kink or antikink, to the 
inhomogeneity $\Psi(\xi,\,\eta)$:
\begin{equation}
\int_{-\infty}^{\infty} u_0'(\xi)\Psi(\xi,\eta)d\xi=0.
\label{solv} \end{equation}
At $w=0$, the well-known result $c=-\kappa$ (curvature driven 
motion \cite{koby}) is evident. The kink propagates along the 
curvature vector in such a way that a convex region occupied by one 
of the alternative equilibria shrinks. The lowest-order general 
equation of motion for a curved kink is written as
\begin{equation}
c(\eta)=-\kappa(\eta) \mp b w(\eta), 
\label{eqmot} \end{equation}
where
\begin{equation}
b = \int_{-\infty}^{\infty} u_0'(\xi)  f_1(u_0(\xi))\, d\xi \left/ 
     \int_{-\infty}^{\infty} [u_0'(\xi)]^2d\xi. \right.
\label{b1} \end{equation}
In particular, for the model Eq.(\ref{sueq}), $u_{0}'(\xi)= 2^{-1/2} 
\mbox{sech}^{2}(\xi/\sqrt{2}), \; f_1(u)=1, \; w=U-v$, and 
$b=3/\sqrt{2}$. The upper sign in Eq.(\ref{eqmot}) applies to a kink 
with the lower state prevailing at $\xi<0$, and the lower, to an 
antikink with the reverse orientation. 

The equation of motion Eq.(\ref{eqmot}) is closed by defining the 
instantaneous position of the kink in a suitable global coordinate 
frame and computing the local curvature. The outer equation 
Eq.(\ref{veq}) can be then solved by substituting the quasistationary 
values $u=u^\pm$ in the regions separated by the kink, and using 
Eq.(\ref{eqmot}) as a kinematic boundary condition at  the kink. 

\section{Regular structures\label{SREG}}

\subsection{Solitary straight kink\label{Sstat1}}

The simplest structure in a system with separated scales is a steadily 
propagating solitary straight kink. Using in Eq.(\ref{oveq}) the 
quasistationary values of the short-range variable $u=u^\pm(v)$, 
which can be computed in the leading order in $\epsilon$ using the 
roots $u_\pm$ of $f_0(u)$, and transforming to the moving frame we 
rewrite it as
\begin{equation}
 \epsilon^2 \gamma^{-1}c_0 v_x  + v_{xx} - v \pm \mu - \nu = 0, 
\label{ovseq} \end{equation}
where $\mu=\frac{1}{2}[g_0(u^+) -g_0(u^-) ]$, 
$\nu=-\frac{1}{2}[g_0(u^+) -g_0(u^-) ]$; the same form is obtained 
directly when the model equation (\ref{sveq}) is used. The small 
parameter in the first term appears when the propagation velocity is 
scaled as specified in the preceding Section. Provided $\gamma \gg 
\epsilon^2$, the propagative term is small, and will be further 
omitted. Then the leading-order unperturbed solution satisfying the 
continuity and smoothness conditions at the 
kink is
\begin{equation}
v= \left\{ \begin{array}{lcc}
\mu [e^{x} - 1] - \nu &\mbox{ at } & x<0 , \\
\mu [1 - e^{-x}] - \nu &\mbox{ at } & x>0. 
     \end{array} \right.
\label{svsoln} \end{equation}
When the short-scale equation (\ref{sueq}) applies, the stationary 
propagation speed, defined by Eq.~(\ref{eqmot})  is $c = -b[U-v(0)]= -
b (U+\nu)$. The kink is at rest when the fraction of the surface 
occupied by the lower state, $\alpha$, is $\alpha=(\epsilon/2\beta) 
(U_0+\nu)$. Since this ratio is small at $\beta=O(1)$, it is likely that 
boundary effects have to be taken into account in this case.

\subsection{Striped pattern\label{Smany}}

A regular striped pattern in the infinite plane with the period $2L= 
2(a_+ +a_-)$ is formed by alternating straight-line kinks at $x=2nL$ 
and antikinks at $x=2(nL+a_+)$. The leading-order stationary solution 
for the long-range variable is
\begin{eqnarray}
v= \mu \left[ 2\cosh\left( x-2nL + a_- \right) \:\frac{\sinh a_+}
         {\sinh L} - 1 \right] - \nu 
 &\mbox{ at }& 2(nL-a_-) <x< 2nL , \nonumber \\
v = \mu \left[ 1- 2\cosh\left( x-2nL - a_+ \right) \:
   \frac{\sinh a_-}
         {\sinh L}  \right] - \nu 
&\mbox{ at } & 2nL<x< 2(nL + a_+)  . 
   \label{mvsoln} \end{eqnarray}
The equilibrium condition, that fixes the relation between the lengths 
$a_\pm$, is $v(0)=v(2a_+)=U$. A short computation yields
 \begin{equation}
U+\nu = \mu \sinh (a_+ - a_-)/\sinh (a_+ + a_-) . 
\label{mstat} \end{equation}
Using also Eq.~(\ref{sgleq}) we obtain 
\begin{equation}
U_0 +\nu = \mu \frac{\sinh (a_+ - a_-)}{\sinh (a_+ + a_-)}
       +\frac{2\beta a_-}{\epsilon(a_+ + a_-)}.
\label{mstatu} \end{equation}
Evidently, $a_+ \gg a_-$, unless $\beta \le O(\epsilon)$.

\subsection{Solitary band\label{Sblob1}}

\begin{figure}
\psfig{figure=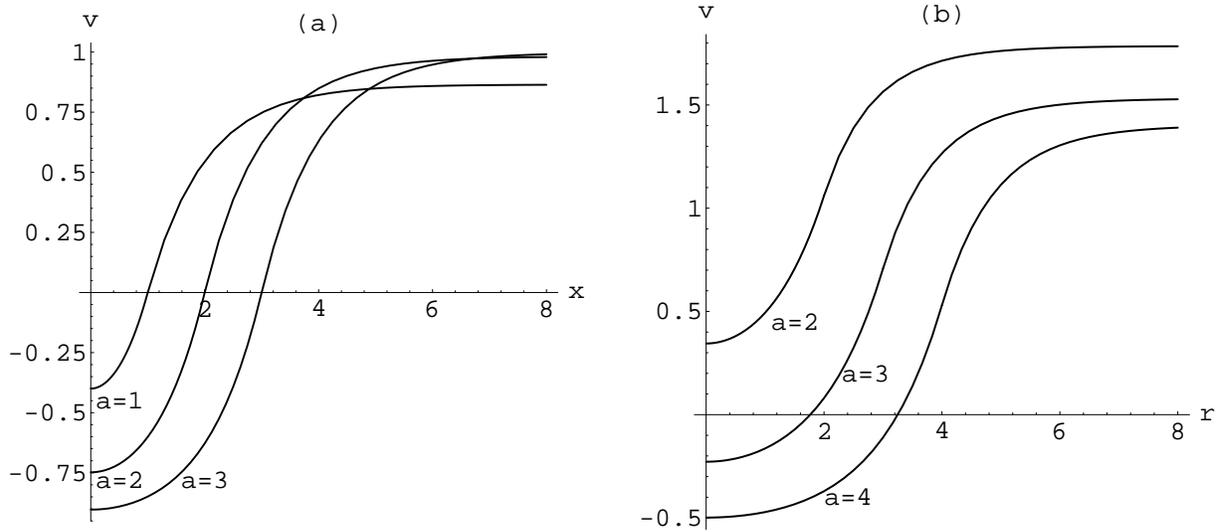}
\caption{The profile of the long-range variable $v$ for a solitary 
band (a) and disk (b) solutions. The value of $v$ is normalized by 
$\mu$, and the value of the parameter $\nu$ is chosen in such a way 
that the bounding kink is stationary in the absence of global control. 
The curves are marked by the values of the band half-width or the 
disk radius $a$.}
 \label{fprofile}
\end{figure}

A solution in the form of a solitary band with $u<0$ immersed in a 
continuum with $u>0$ can be obtained from Eq.~(\ref{mvsoln}) in the 
limit $a_+ \to \infty$. The solution, that satisfies the continuity and 
smoothness conditions at the kink and the antikink located at 
$x=\pm a$, is
\begin{equation}
v= \left\{ \begin{array}{lcc}
\mu (2e^{-a}\cosh x -1) - \nu   &\mbox{ at } & |x|<a , \\
\mu (1 - 2 e^{-|x|}\sinh a) - \nu  &\mbox{ at } & |x|>a. 
      \end{array} \right.
\label{1soln} \end{equation}
Some typical profiles of $v$ are shown in Fig.~\ref{fprofile}a. The 
stationarity condition is
\begin{equation}
\nu + U = \mu e^{-2a}.
\label{1stat} \end{equation}
Solitary structures naturally appear at $\beta =O(1)$. Since the 
interaction between distant kinks is exponentially small, one can also 
envisage a pattern consisting of $N$ well separated parallel bands. 
When such a structure is placed on a plate with the width 
$L/\epsilon$, the fraction of the surface occupied by the lower state 
is $\alpha=2\epsilon Na/L$. Then $U=U_0-2\beta_1 a$, where 
$\beta_1 = 2N\beta a/L$. It is easy to see that $U_0$ can be 
compensated by a shift of $\nu$; further on, we shall set it therefore 
to zero. Then the stationarity condition Eq.~(\ref{1stat}) can be 
written in the form
\begin{equation}
\nu - 2\beta_1 a  =  \mu e^{-2a} ,
\label{1statu} \end{equation}
If this equation is solved graphically, stable solutions can be obtained 
only when the {\em absolute value} of slope of the straight line 
representing the left-hand side is {\em less} than that of the 
right-hand side at the intersection point. The necessary condition for 
the existence of such solutions is $\beta_1 \le \mu$. Within the 
parametric range $\mu > \beta_1 > 0 $, $\mu > \nu > \nu_c \equiv 
\beta_1 [1- \ln (\beta_1/\mu)]$, there are two solutions, of which 
one corresponding to a smaller width is stable according to the above 
criterion. The largest possible width, achieved at $\nu=\nu_c$, is 
$a=\frac{1}{2} \ln (\mu/\beta_1)$. 

\subsection{Solitary disk\label{Sblob2}}

In 2d, the kink should acquire a circular shape due to the ``line 
tension'', and the stationary solution depends only on the long-range 
radial coordinate $r$. The stationary outer equation is
\begin{equation}
v_{rr}+r^{-1}v_r - v - \nu \pm \mu= 0,
\label{a2veq} \end{equation}
The stationary solution for a disk with $u<0$ immersed in a 
continuum with $u>0$, satisfying the continuity and smoothness 
conditions at the kink located at $r=a$, is
 \begin{equation}
v= \left\{ \begin{array}{lcc}
\mu [2aK_1(a)I_0(r) - 1] - \nu &\mbox{ at } & r<a ,\\
\mu [1 - 2aI_1(a)K_0(r)] - \nu &\mbox{ at } & r>a, 
    \end{array} \right.
\label{2soln} \end{equation}
where $I_n,\:K_n$ are modified Bessel functions. The identity 
$K_1(a)I_0(a) + K_0(a)I_1(a)=a^{-1}$ should be used when the 
boundary conditions are checked. Some profiles of $v$ are shown in 
Fig.~\ref{fprofile}b. One can notice that the value of the long-range 
variable at the kink location is shifted upwards to 
compensate the curvature-driven shrinking action; this shift grows 
with the decreasing disk radius. 

\begin{figure}
\psfig{figure=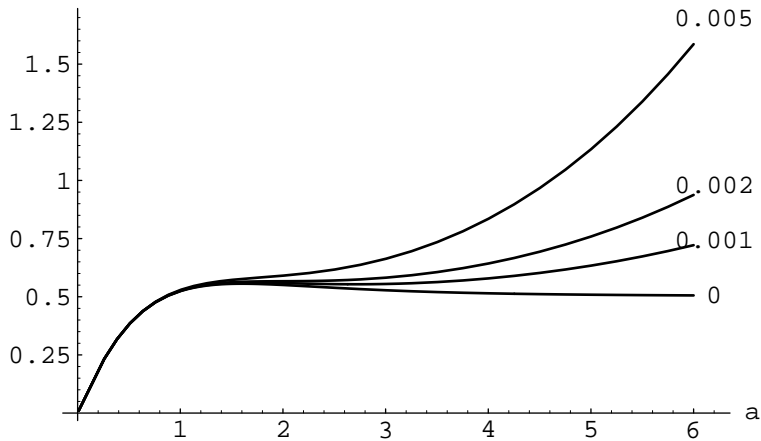}
\caption{Construction of stationary solitary disk solutions. Each curve 
presents the plot of the r.h.s. of Eq.~(\protect\ref{2stat}), and is 
marked by the appropriate value of the global control parameter 
$\beta_2$. Solutions are obtained as intersection points with straight 
lines presenting the l.h.s. of Eq.~(\protect\ref{2stat}).}
 \label{fdstat}
\end{figure}

Using $v(a)$ and $\kappa= a^{-1}$ in Eq.(\ref{eqmot}) yields the 
stationarity condition 
\begin{equation}
U+\nu= - (b a)^{-1} +
    \mu a\left[ K_1(a)\,I_0(a) - K_0(a)\,I_1(a) \right] .
\label{cstat} \end{equation}
If a pattern consisting of $N$ well separated disks is formed on a 
plate with the surface area $(L/\epsilon)^2$, the fraction of the 
surface occupied by the lower state is $\alpha=\epsilon \pi N 
(a/L)^2$. Then $U=U_0-2\pi \beta N(a/L)^2$, (where $U_0$ can be 
again set to zero), and the stationarity condition Eq.~(\ref{cstat}) can 
be written, using the parameter $\beta_2= \pi \epsilon N\beta/ L^2$, 
in the form
\begin{equation}
\nu a + b^{-1} =  2\beta_2 a^3 + 
    \mu a^2 \left[ K_1(a)\,I_0(a) - K_0(a)\,I_1(a) \right] .
\label{2stat} \end{equation}
If this equation is solved graphically (Fig.~\ref{fdstat}), the slope of 
the left-hand side of this equation at the intersection point 
corresponding to a stable solution should be {\em algebraically 
larger} than the slope of the right-hand side. When it is so, the 
effective repulsive action due to the long-range field decreases with 
growing radius faster than the shrinking action due to the line 
tension. Whenever Eq.(\ref{cstat}) has a single solution, this solution 
should be unstable. When an additional pair of solutions bifurcates at 
a tangent point, one of these solutions is stable according to the 
above criterion. Further stability analysis is found in the next 
Section. 


\section{Stability of regular structures\label{SSTAB}}

The ``naive'' stability criteria in the preceding Section refer only to 
the stability to perturbations that do not reduce the symmetry of the 
problem. In this Section, we shall investigate stability to asymmetric 
perturbations. Perturbations of this kind do not change the fraction 
of the surface occupied by either state, and therefore do not involve 
the global control variable. Throughout this Section, we shall assume 
that $\gamma$ is not exceedingly small, i.e. $\gamma \ge 
O(\epsilon)$. Under this condition, the distribution of the long-range 
variable follows slow shifts of the kink position quasistationarily, 
unless at distances that are large even on the extended scale. 

\subsection{Stability of a solitary straight kink\label{Sstab1}}

We shall now apply the equation of motion Eq.(\ref{eqmot}) to the 
problem of stability of a solitary straight kink to infinitesimal 
perturbations. The suitable global coordinate frame for this problem 
is a Cartesian frame $(x,y)$ comoving with the unperturbed kink. 
Denoting the instantaneous position of the kink relative to its 
unperturbed position as $x =\zeta(y,t)$, we take note that, as long as 
the amplitude of the perturbation is much smaller than its 
wavelength, the normal vector, defining the direction of the kink 
propagation, is almost parallel to the $x$ axis, and the propagation 
speed is expressed as $\zeta_t=\epsilon^2 \gamma^{-1} c$. In the 
same approximation, the curvature is given by $\kappa =-
\zeta_{yy}$. Expanding also the long-range variable in the vicinity of 
the unperturbed position, we reduce 
Eq.(\ref{eqmot}) to the form
\begin{equation}
\zeta_\tau = \zeta_{yy} + 
          b \left[V(0,y,\tau) + v'_s(0) \zeta \right] .
\label{smot} \end{equation}
Here $v'_s(0)=\mu$ is the stationary flux at the kink following from 
Eq.(\ref{svsoln}), $V$ is the perturbation of the long-range variable, 
and $\tau= \epsilon^2 \gamma^{-1} t$ is the rescaled time variable.

The correction to $v$ due to a weak perturbation of the kink can be 
computed in a simple way by observing that any shift of the kink 
position by an infinitesimal increment $\zeta$ is equivalent to 
switching the sign of the source term in Eq.(\ref{sveq}) in a narrow 
region near $x=0$. This approach is akin to the ``singular limit 
eigenvalue'' method applied earlier to the stability analysis of kinks 
in one dimension \cite{nish}. Although the perturbation $\xi$ is 
habitually presumed to be arbitrarily small for the purpose of linear 
stability analysis, the method is actually applicable also to finite 
perturbations of the kink position, provided they are small on the 
extended scale, i.e. $\zeta \le O(\epsilon)$. 

At $\gamma \ge O(\epsilon)$, the quasistationary form of the 
perturbation equation is applicable: 
\begin{equation}
          \nabla^2 V -  V = 2 \mu \zeta(y,t) \delta (x).
      \label{sVeq}     \end{equation}

In a standard way, we are looking for a solution in a spectral form
\begin{equation}
 \zeta(y,\tau) = \chi(k, \lambda) e^{\lambda \tau + iky} , \;\;\; 
   V (x,y,\tau)= \psi(x; k, \lambda) e^{\lambda \tau + iky}.       
\label{fourier}     \end{equation}
The dispersion relation $\lambda(k^2)$ determines the stability to 
infinitesimal perturbations. Using Eq.~(\ref{fourier}) in 
Eq.(\ref{sVeq}) yields
\begin{equation}
  \psi_{xx} - q^2\psi = 2\mu \zeta(y,t) \delta (x),
      \label{spsieq}     \end{equation}
where $q^2=1 +k^2$. The solution of this equation is
\begin{equation}
\psi(x) = -\mu\chi(k, \lambda) q^{-1}  e^{-q|x| }.
 \label{sVsoln} \end{equation}
Using Eqs.(\ref{svsoln}), (\ref{fourier}), (\ref{sVsoln}) in 
Eq.(\ref{smot}) 
yields the dispersion relation 
\begin{equation}
\lambda = - k^2 + b \mu (1 - q ^{-1}) .
  \label{disp0} \end{equation}

The eigenvalues $\lambda(k^2)$ are always real, and, as expected, 
$\lambda$ vanishes at $k=0$, which reflects the translational 
symmetry of the kink. The loss of stability to long-scale 
perturbations occurs when the derivative $d\lambda /d(k^2)$ at 
$k=0$ becomes positive. This happens at $b \mu >2$. Since the 
function $\lambda (k^2)$ is convex, the loss of stability always occurs 
with growing $\mu$ in the long-scale mode.

\subsection{Stability of a solitary band\label{Sstab2}}

The equations of motion for infinitesimal displacements 
$\zeta^{(1)},\: \zeta^{(2)}$ of the kink at $x=a$ and the antikink at 
$x=-a$ that confine a solitary band (Section \ref{Sblob1}) are written 
analogous to Eq.(\ref{smot}). We take note that opposite signs appear 
in the equations of motion Eq.~(\ref{eqmot}) for the kink and the 
antikink. Using the expression for the stationary fluxes following 
from Eq.(\ref{1soln}), we obtain
\begin{equation}
\zeta^{(j)}_\tau = \zeta^{(j)}_{yy} + 
          b \left[ \pm V (\pm a, y,\tau) +  
         \mu (1 - e^{-2a}) \zeta^{(j)} \right] . 
\label{1mot} \end{equation}
It is advantageous to consider the symmetric and antisymmetric 
combinations of the displacements, $\zeta^\pm = \frac{1}{2} 
(\zeta^{(1)} \pm \zeta^{(2)})$, that may lead, respectively, to {\em 
zigzag} or {\em varicose} instabilities. It is clear from Eq.(\ref{1mot}) 
that $\zeta^\pm$ are coupled, respectively, to the antisymmetric and 
symmetric parts $V^\mp$ of the perturbation field $V$: 
\begin{equation}
\zeta^\pm_\tau = \zeta^\pm_{yy} + 
       b \left[ V^\mp( a, y,\tau) +  \mu (1 - e^{-2a}) \zeta^\pm \right]. 
\label{2mot} \end{equation}

The quasistationary  equations for both perturbation fields, $V^\mp$, 
are identical: 
\begin{equation}
 \nabla^2 V^\mp -  V^\mp = 2 \mu \zeta^\pm(y,\tau) \delta (x-a) .
      \label{2Veq}     \end{equation}
Setting as before $V^\pm = \psi^\pm e^{\lambda \tau + iky}$, 
$\zeta^\pm = \chi^\pm e^{\lambda \tau + iky}$ yields
\begin{eqnarray}
\psi^+(x) &=& \left\{ \begin{array}{lcc}
-2\mu\chi^+ q ^{-1} e^{-qa }\cosh qx  
           &\mbox{ at } & |x|<a ,\\
-2\mu\chi^+ q ^{-1} e^{-q|x| }\cosh qa 
           &\mbox{ at } & |x|>a ,     \end{array} \right. \\
\psi^-(x) &=& \left\{ \begin{array}{lcc}
-2\mu\chi^- q ^{-1} e^{-qa }\sinh qx 
           &\mbox{ at } & |x|<a ,\\
-2 \mbox{ sign}(x)\,\mu\chi^- q ^{-1} e^{-q|x| }\sinh qa 
           &\mbox{ at } & |x|>a .    \end{array} \right.
\label{p2soln} \end{eqnarray}
 
The zigzag and varicose branches of the dispersion relation are given 
by
\begin{eqnarray}
\lambda^+ &=& - k^2 + b\mu \left[1 - e^{-2a } - q ^{-1}
     \left(1 - e^{-2qa }\right)\right],         \nonumber \\
 \lambda^- &=& - k^2 + b\mu \left[1 - e^{-2a } - q ^{-1}
     \left(1 + e^{-2qa }\right)\right].
   \label{2disp} \end{eqnarray}

Vanishing $\lambda^-$ at $k=0$ reflects the translational symmetry 
of the band. Due to convexity of the function $\lambda^-(k^2)$, the 
loss of stability first occurs at $k=0$. The solution is unstable when 
the derivative $d\lambda^-/dk$ at $k=0$ is positive. The critical 
point of the long-scale zigzag instability is
\begin{equation}
\mu = \frac{2/b} {(1 +2a) e^{-2qa }-1}.
\label{2alimit} \end{equation} 
The varicose branch, $\lambda^+(k^2)$ lies persistently lower than 
the zigzag one. Thus, the stability limit is always given by 
Eq.(\ref{2alimit}). The most dangerous perturbation corresponds to 
the long-wave limit $k\to 0$, i.e. $q\to 1$. The limit of the zigzag 
instability is shifted to higher values of $\mu$ with a decreasing 
band thickness. 

\subsection{Stability of a solitary disk\label{Sstabd}}

In order to check stability of the stationary solution that has the 
form of a solitary disk with the radius $a$, as in Section \ref{Sblob2}, 
we use the cylindrical coordinates $(r,\phi)$, and set $r =a[1+ \zeta 
(\phi,t)]$, where $\zeta (\phi,t) \ll 1$. The curvature of the kink is 
expressed as $\kappa =a^{-1}(1- \zeta - \zeta_{\phi\phi})$. 
Expanding, as before, the long-range variable in the vicinity of the 
unperturbed position brings Eq.(\ref{eqmot}) to the form
\begin{equation}
\zeta_\tau = a^{-2} \left( \zeta_{\phi\phi} + \zeta \right) +
          b \left[ a^{-1}V(0,\phi,\tau) +  v'_s(0) \zeta \right] ,
\label{smotd} \end{equation}
where $v'_s(0)=2\mu aI_1(a)K_1(a)$ is the stationary flux at the 
kink following from Eq.(\ref{2soln}). The quasistationary equation 
for the perturbation field $V$ is 
\begin{equation}
          \nabla^2 V -  V = 2 \mu a\zeta(\phi,t) \delta (r-a).
      \label{sVeqd}     \end{equation}
We are looking for a solution in the form
\begin{equation}
 \zeta(\phi,\tau) = \chi_n e^{\lambda_n \tau + in\phi} , \;\;\; 
   V (r,\phi,\tau)= \psi(r; n) e^{\lambda_n \tau + in\phi},       
\label{fourierd}     \end{equation}
where $n$ is an integer. The equation of $\psi$ is
\begin{equation}
  \psi_{rr} + \frac{1}{r}\psi_r - \left(1 + \frac{n^2}{r^2}\right)\psi = 
        2 \mu a \chi_n  \delta (r-a).
      \label{spsieqd}     \end{equation}
The solution of this equation is
\begin{equation}
\psi(r)= \left\{ \begin{array}{lcc}
- 2\mu a^2 K_n(a)I_n(r)  &\mbox{ at } & r<a ,\\
- 2\mu a^2 I_n(a)K_n(r)  &\mbox{ at } & r>a. 
      \end{array} \right.
\label{sVsolnd} \end{equation}
Using Eqs.(\ref{2soln}), (\ref{fourierd}), (\ref{sVsolnd}) in 
Eq.(\ref{smot}) yields the dispersion relation 
\begin{equation}
\lambda_n = - (n^2-1)/a^2 + b\mu a [I_1(a)K_1(a)-I_n(a)K_n(a)] .
  \label{disp0d} \end{equation}

As expected, $\lambda_n$ vanishes at $n=1$, which corresponds to a 
shift of the disk without deformation. One can check numerically 
that, of all angular modes, the quadrupole mode $n=2$, deforming 
the disk into an ellipse, is the most dangerous one. At large $a$, this 
can be seen by using the asymptotics of Bessel functions to compute 
the asymptotic value of $\mu$ at the instability threshold
\begin{equation}
\mu_c(n) = \frac{(n^2-1)/b}{  \frac{1}{4}
  \left[ \Gamma\left(\frac{3}{2} + n\right)/ 
   \Gamma\left(-\frac{1}{2} + n\right) \right] ^2 -\frac{9}{64}  },
\label{asymstab}  \end{equation}
where $\Gamma(x)$ is the gamma-function. The asymptotic limiting 
value increases with $n$ at $n > 2$.

\subsection{Stability of a striped pattern\label{Sstabmany}}

A quasistationary perturbation $V$ of the long-range field in 
alternating regions of a striped pattern with the period $2L$ (Section 
\ref{Smany}) due to displacements of respective kinks at $x=2nL$ by 
$\zeta^{(1,n)}$ and antikinks at $x=2(nL+a_+)$ by $\zeta^{(2,n)}$ 
obeys the equation 
\begin{equation}
     \nabla^2 V - V = 2 \mu \sum_n (-1)^n \left[
    \zeta^{(1,n)}\delta (x-2nL) - \zeta^{(2,n)}\delta (x-2nL-2a_+) 
\right].
      \label{mVeq}     \end{equation}
The equation of motion of the kinks is obtained, analogous to 
Eq.(\ref{smot}), by using in Eq.(\ref{eqmot}) $v=v'_s(2nL)\zeta^{(n)} 
+V(2nL,y,\tau)$, where $v'_s(2nL) \equiv J=2\mu \sinh a_+ \sinh a_-
/\sinh L$ is the stationary flux following from Eq.(\ref{mvsoln}); at 
the antikink locations, $x=2(nL+a_+)$, $v'_s(2nL)=-J$, and the 
respective values  enter the equation of motion with the reverse 
sign:
\begin{eqnarray}
\zeta^{(1,n)}_\tau &=& \zeta^{(1,n)}_{yy} + 
          b \left[J\zeta^{(1,n)} +  V( 2nL,y,\tau) \right] ,  \nonumber \\
\zeta^{(2,n)}_\tau &=& \zeta^{(2,n)}_{yy} + 
         b \left[J\zeta^{(2,n)} -  V( 2nL+2a_+,y,\tau) \right] .
\label{mmot} \end{eqnarray}
Presenting the perturbation $V$ in each successive stripe with a 
prevailing positive or negative state of the short-scale variable as 
$V^{(j,n)} = \psi^{(j,n)} e^{\lambda \tau + iky}$, and kink 
displacements as $\zeta^{(j,n)}= \chi^{(j,n)} e^{\lambda \tau + iky},\; 
j=1,2$, we write the equation of the spectral components  
\begin{equation}
  \psi^{(j,n)}_{xx} - q^2 \psi^{(j,n)} =  0.
      \label{mpsieq}     \end{equation}
The matching conditions, taking into account localized perturbations 
due to displacements of respective kinks and antikinks, are 
\begin{equation}    \begin{array}{lll} 
\psi^{(1,n)} = \psi^{(2,n)} , & 
 \psi^{(1,n)}_x - \psi^{(2,n)}_x
    = 2 \mu \chi^{(1,n)} & \mbox{  at  } x=2nL, \\
 \psi^{(1,n)} = \psi^{(2,n+1)} , & 
  \psi^{(1,n)}_x - \psi^{(2,n+1)}_x
    = 2 \mu \chi^{(2,n)} & \mbox{  at  } x=2(nL+a_+).
   \end{array}   \label{mmatch}     \end{equation}
The solution of Eq.(\ref{mpsieq}) is expressed as 
\begin{equation}
\psi^{(j,n)} = A^{(j,n)} e^{qx} + B^{(j,n)} e^{-qx},
      \label{psin}     \end{equation}
We shall define the coefficients in the above expression, as well as 
the displacements at particular locations, through respective 
generating functions:
\begin{eqnarray}
  \alpha_j( z)  = \sum_{n=-\infty}^{\infty} A^{(j,n)} e^{(iz+2qL)n }, 
& \;\;\;&
 \beta_j( z) =  \sum_{n=-\infty}^{\infty} B^{(j,n)} e^{(iz-2qL)n }, 
  \nonumber \\ 
\gamma_j( z)  = \sum_{n=-\infty}^{\infty}\chi^{(j,n)} e^{izn} ,
& \;\;\;& j=1,2.  
 \label{mfourier}     \end{eqnarray}
Multiplying the matching conditions Eq.(\ref{mmatch}) by $e^{izn}$ 
and summing over $n$ yields a set of equations for $\alpha_j( z), \: 
\beta_j(z)$: 
\begin{eqnarray}   
 \alpha_1( z) - \alpha_2( z) + \beta_1( z) - \beta_2( z) & = & 0, 
\nonumber \\
  \alpha_1( z) - \alpha_2( z) -  \beta_1( z) + \beta_2( z) 
         & = & 2 \mu q^{-1}\gamma_1(z), \nonumber \\
 e^{2qa_+}\alpha_1( z) - e^{-(iz +2qa_-)}\alpha_2( z) + 
         e^{-2qa_+}\beta_1( z) - e^{-(iz-2qa_-)}\beta_2( z) & = & 0, 
\nonumber \\
  e^{2qa_+}\alpha_1( z) - e^{-(iz +2qa_-)}\alpha_2( z) - 
         e^{-2qa_+}\beta_1( z) + e^{-(iz-2qa_-)}\beta_2( z) 
     &=& 2 \mu q^{-1}\gamma_2(z).
      \label{zmatch}     \end{eqnarray}
Two more equations are obtained in the same way from 
Eq.(\ref{mmot}):
\begin{eqnarray}
(\lambda + k^2 - bJ)\gamma_1(z) & = & 
      \alpha_1( z) + \beta_1( z) , 
      \nonumber \\
(\lambda + k^2 - bJ)\gamma_2(z) & = & 
   - e^{2qa_+} \alpha_1( z)  - e^{-2qa_+}\beta_1( z) .   
\label{zmot} \end{eqnarray}

The general dispersion relation, obtained as the condition that 
Eqs.~(\ref{zmatch}), (\ref{zmot}) have a non-trivial solution, is quite 
cumbersome. As in Section \ref{Sstab2}, it is convenient to work 
with zigzag and varicose modes, that correspond, respectively, to 
like and opposite signs of displacements of adjacent kinks and 
antikinks. These modes, obtained by introducing the symmetric and 
antisymmetric combinations $\gamma^\pm= 
\frac{1}{2}(\gamma_1\pm\gamma_2)$, decouple, however, only at 
$z=0$. The symmetric (zigzag) and antisymmetric (varicose) branches 
of the dispersion relation at $z=0$ are computed as
\begin{eqnarray}
\lambda^+ &=& - k^2 + 2b\mu \left(
        \frac{\sinh a_+ \sinh a_-}{\sinh L}  -
   \frac{\sinh qa_+ \sinh qa_-}{q\sinh qL}\right) , 
        \nonumber \\
 \lambda^- &=& - k^2 + 2b\mu \left( 
     \frac{\sinh a_+ \sinh a_-}{\sinh L}  -
   \frac{\cosh qa_+ \cosh qa_-}{q\sinh qL}\right) .
      \label{mdisp} \end{eqnarray}

The varicose branch $\lambda^-(k^2)$ lies persistently lower than 
$\lambda^+(k^2)$, and therefore can be disregarded. Vanishing 
$\lambda^+$ at $k=0$ reflects the translational symmetry of the 
entire pattern. The loss of stability in the long-scale zigzag mode 
occurs when the derivative $\partial \lambda^+/\partial (k^2)$ at 
$k=0$ vanishes. 

\begin{figure}
\psfig{figure=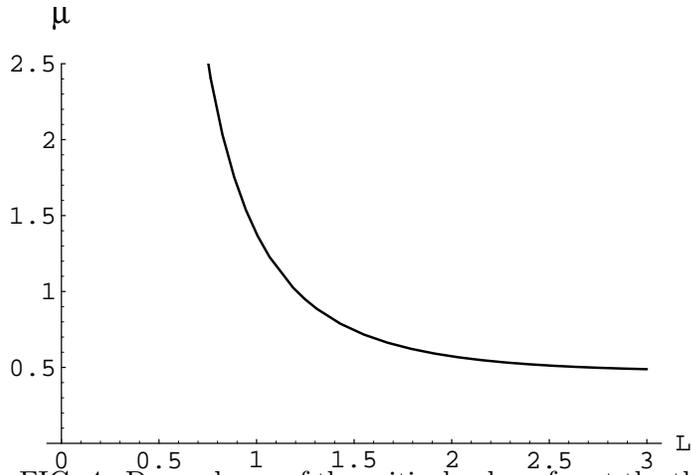}
\caption{Dependence of the critical value of $\mu$ at the threshold 
of the long-scale concerted zigzag instability of a symmetric striped 
pattern on the half-period $L$.} \label{Fmua}
\end{figure}

A simpler form of the dispersion relation is obtained at $\nu=0$ 
when the striped pattern is symmetric, with $a^+=a^-=L$. The zigzag 
branch of the dispersion relation is
\begin{equation}
\lambda = - k^2 + b\mu\left[ \tanh L
      - \frac{q^{-1}\tanh qL}
       {1-\sin^2 \frac{z}{2}  \left(1-\tanh^2 qL\right)} \right].
 \label{meqdisp} \end{equation}
One can clearly see here that $\lambda$ decreases monotonically 
with growing $z$, so that when $\mu$ increases the loss of stability 
always occurs in a concerted long-scale zigzag mode $z=0$. Again, the 
bracketed expression in Eq.(\ref{meqdisp}) is a convex function of 
$k^2$, and therefore the loss of stability should first occur at $k=0$. 
The instability limit shifts to larger $\mu $ as the wavelength 
decreases, as shown in Fig.~\ref{Fmua}. In the other limit, $a^+ \gg a^-
$, Eq.~(\ref{mdisp}) reduces to the dispersion relation for a solitary 
band,  Eq.~(\ref{2disp}).

\section{Discussion\label{sdiscus}}

\begin{figure}
\psfig{figure=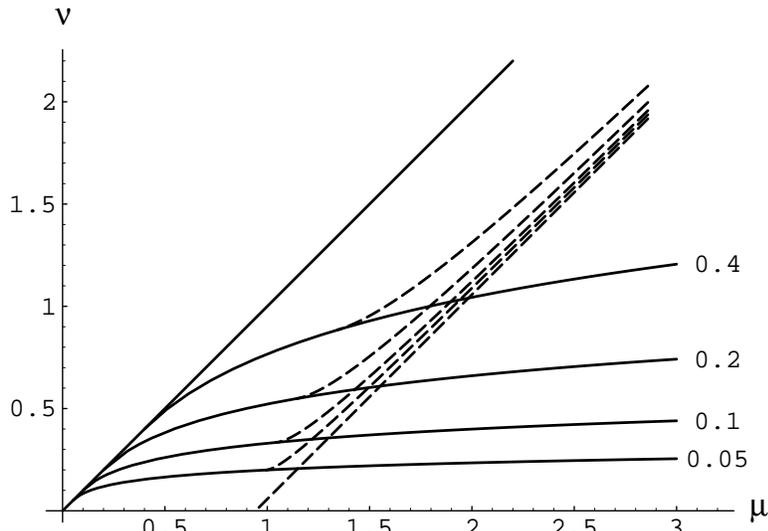}
\caption{Existence and stability regions of the solitary band solution 
in the parametric plane $(\mu,\nu)$. The solution exists in the region 
bounded by the diagonal $\nu=\mu$ and the solid line marked by an 
appropriate value of the global control parameter $\beta_1$. Dashed 
lines denote the limits of the zigzag instability at the same values of 
the parameter $\beta_1$.} \label{fband}
\end{figure}

Although both the zigzag instability of the solitary band solution and 
the quadrupole instability of the solitary disk are not coupled to the 
global control parameter $\beta$, the stability boundary in the 
parametric plane $(\mu,\nu)$ depends on the global parameter as 
well because of the parametric dependence of the stationary solution. 
The existence and stability regions of the solitary band solution are 
shown in Fig.~\ref{fband}. In the absence of global control, the 
solution exists in the entire triangle $\mu>\nu>0$, but suffers zigzag 
instability to the right of the dashed line departing from the $\mu$ 
axis. Both the existence and stability limits shift upwards as the 
parameter $\beta_1$ increases. 

\begin{figure}
\psfig{figure=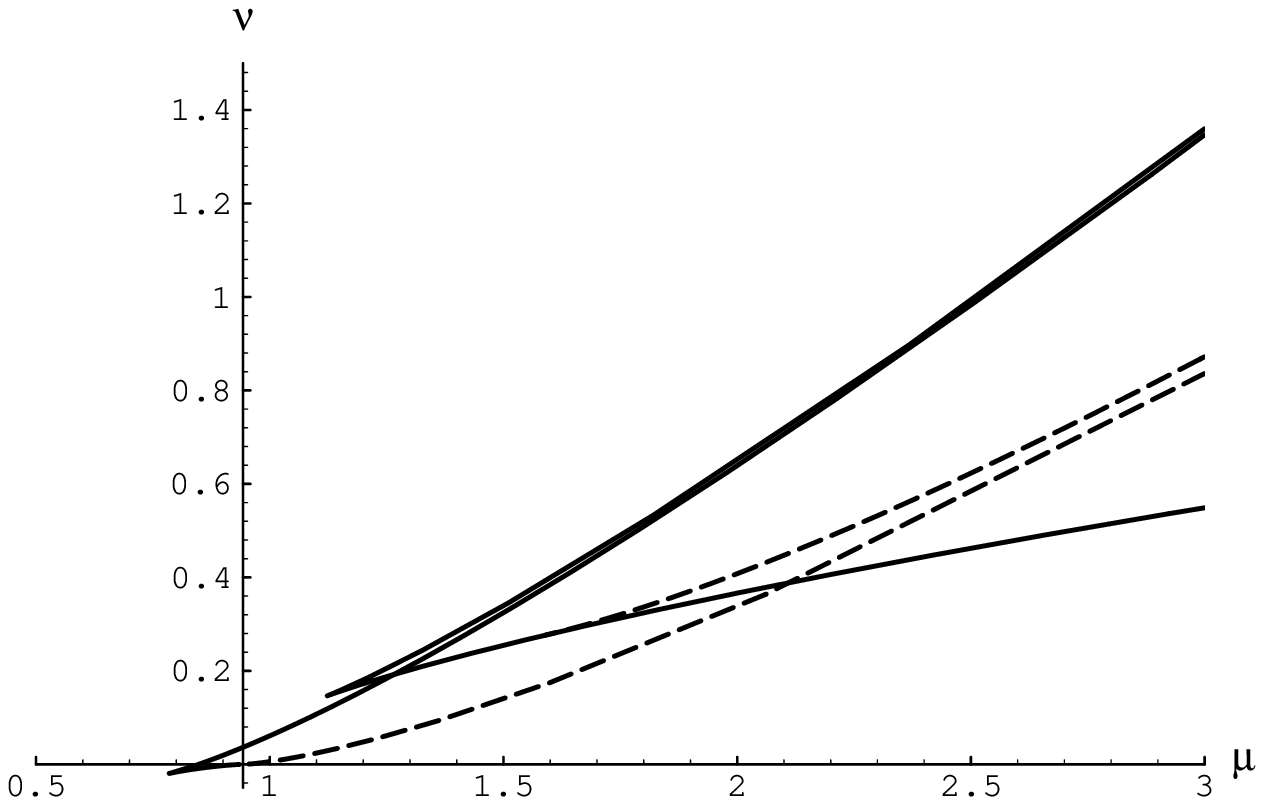}
\psfig{figure=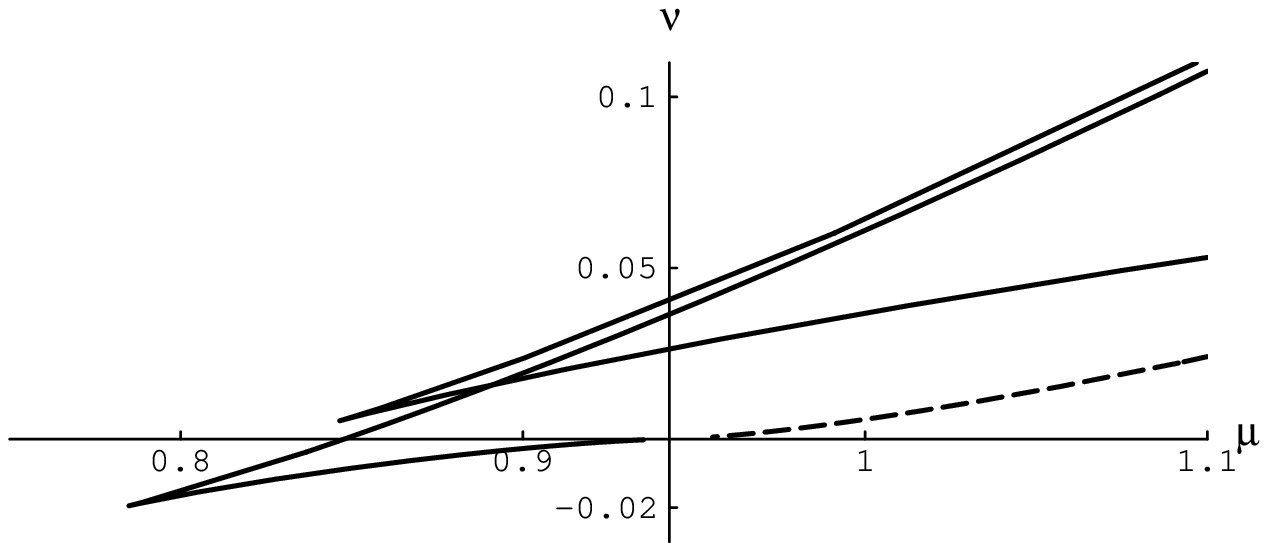}
\caption{Existence and stability regions of the solitary disk solution 
in the parametric plane $(\mu,\nu)$ for the values of the global 
control parameter $\beta_2=0$, $\beta_2=10^{-2}$ (a) and 
$\beta_2=0$, $\beta_2=10^{-3}$ (b). Dashed lines denote the 
instability limits in the quadrupole mode. At $\beta_2= 10^{-3}$, the 
instability limit is outside the depicted region.} \label{fdisk}
\end{figure}

The existence and stability regions of the solitary disk solution  are 
shown in Fig.~\ref{fdisk}. At $\beta_2=0$, the tip of the existence 
region penetrates into the half-plane $\nu<0$; at $\mu>b^{-1}$, the 
lower boundary of the existence region coincides with the axis 
$\nu=0$. Two other cusped curves, drawn for $\beta_2=10^{-2}$ 
(Fig.~\ref{fdisk}a) and $\beta_2=10^{-3}$ (Fig.~\ref{fdisk}b), bear 
witness that the existence conditions are very sensitive to small 
changes of the global parameter $\beta_2$. This sensitivity is, 
indeed, essential, since the definition of the parameter $\beta_2$ 
includes the small parameter $\epsilon$, and therefore $\beta_2$ 
(or, at least, the ratio $\beta_2/\beta_1$) should be small. The limits 
of the quadrupole instability are shown in the same picture by 
dashed lines (at $\beta_2= 10^{-3}$, the instability limit is outside 
the depicted region). 

\begin{figure}
\psfig{figure=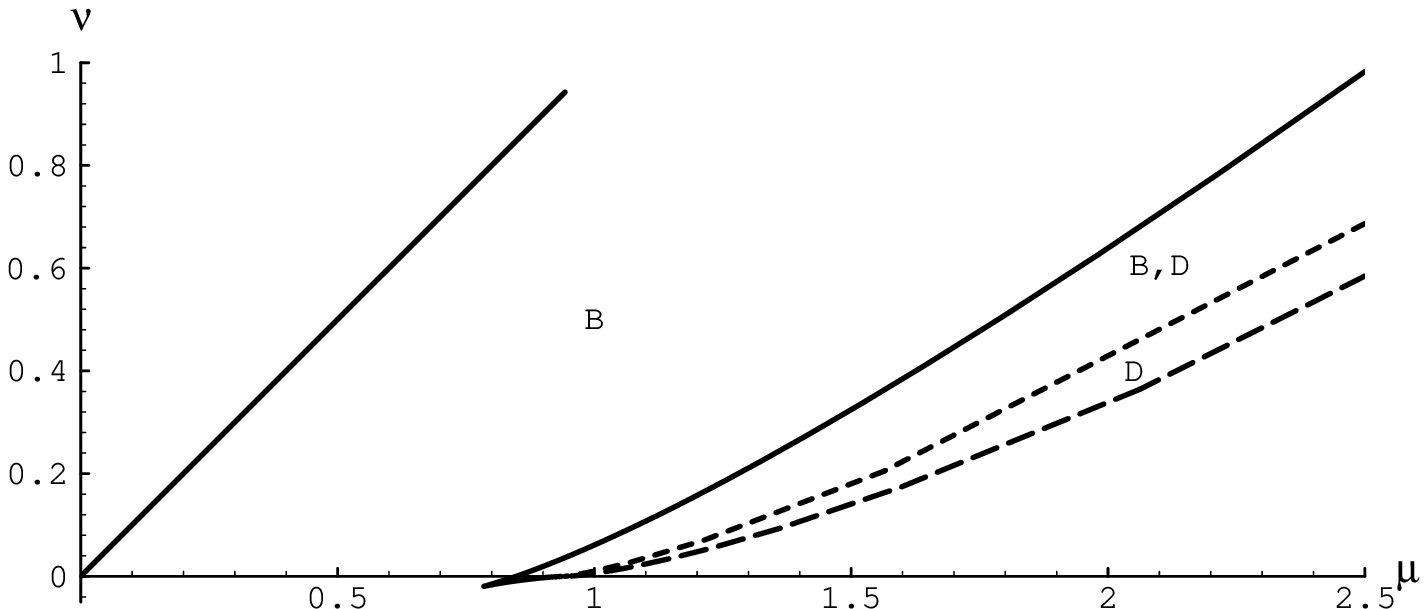}
\psfig{figure=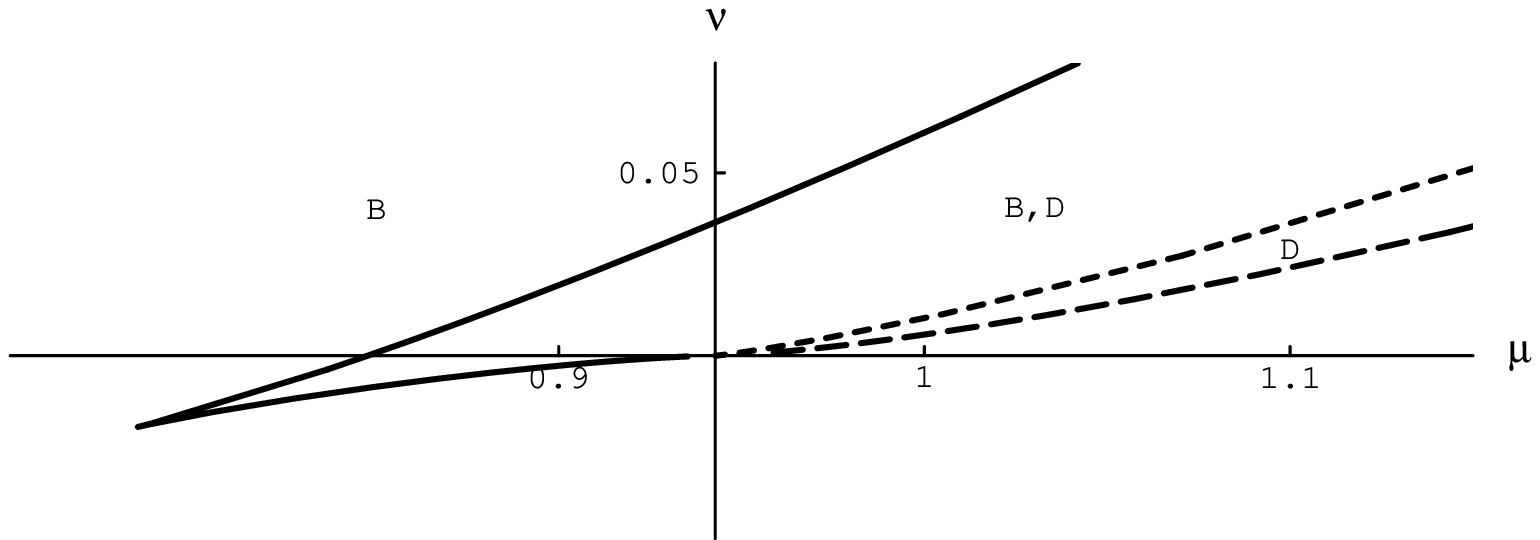}
\psfig{figure=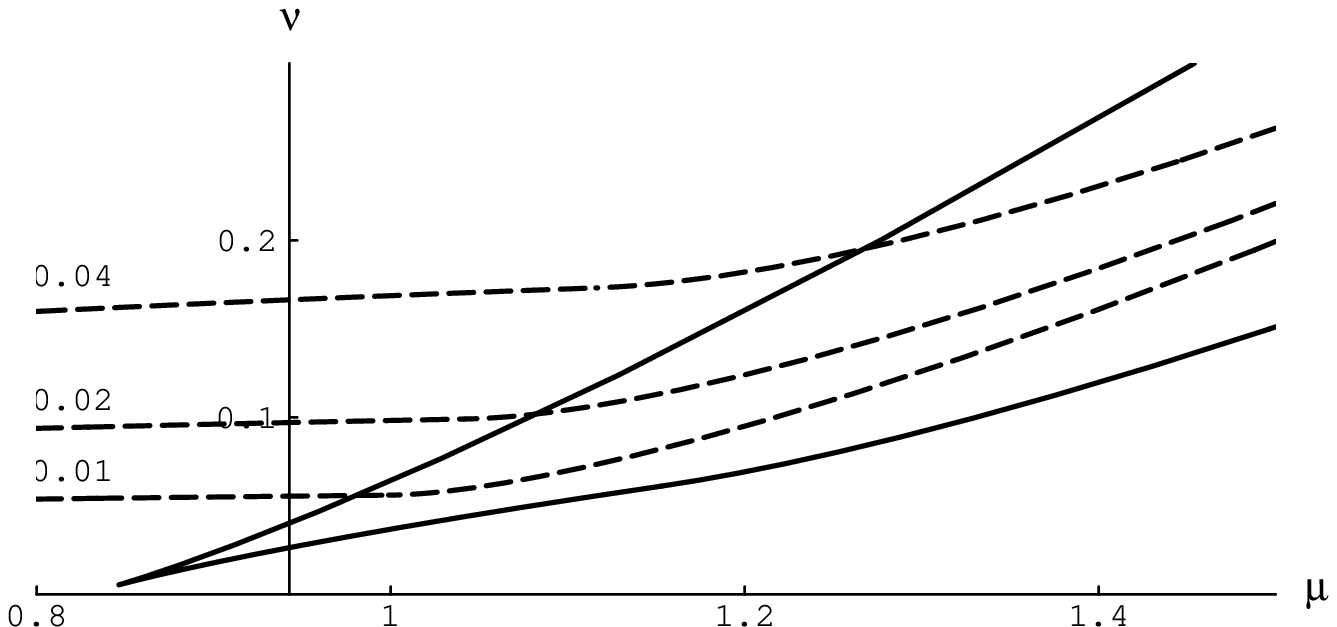}
\caption{Comparison of the existence and stability regions of the 
solitary band (B) and disk (D) solutions in the parametric plane 
$(\mu,\nu)$ for the values of the global control parameter 
$\beta_1=\beta_2=0$ (a,b), and $\beta_2=10^{-3}$ (c). In (a,b), solid 
lines denote the bounds of existence, and dashed lines, the stability 
limits of both the band and the disk solutions. The region adjacent to 
the cusp is shown on a larger scale in (b). The 
solid cusped curve in (c) shows the existence bounds of stable 
solitary disk solutions at $\beta_2=10^{-3}$. The dashed lines in (c) 
denote the existence bounds of stable solitary band solutions, 
and are marked by the appropriate values of $\beta_1$.} 
\label{fbandisk}
\end{figure}

The regions where either or both band and disk solutions can appear 
are shown together in Fig.~\ref{fbandisk}. The coexistence region is 
most extensive when there is no global control 
(Fig.~\ref{fbandisk}a,b). One can recall that coexistent striped and 
dotted patterns (which can be viewed as arrays of band or disk 
structures) were also found experimentally \cite{quy2}. A narrower 
coexistence region is seen in Fig.~\ref{fbandisk}c, where the region of 
{\em stable} solitary disk solutions at $\beta_2=10^{-3}$, bounded 
by the solid cusped line, is shown together with dashed lines 
representing the lower bounds of the existence of {\em stable} 
solitary band solutions at several values of $\beta_1$. The 
coexistence region shrinks with increasing global control very fast, as 
the limit of the zigzag instability of solitary band solutions shifts to 
the left and upwards, and at the same time the tip of the cusped 
region of stable disk solutions retreats to higher values of both 
$\mu$ and $\nu$.

This picture is in qualitative agreement with the results of the 
computer experiment by Middya and Luss \cite{luss93}, although in 
the latter work there was no separation between characteristic scales 
of both local variables. In runs without global control, Middya and 
Luss observed, depending on initial conditions, either bands or disks. 
Although typically a large number of either objects were present, 
they could be seen, essentially, as solitary structures, since the 
diffusivity of both local variables was small, and therefore separated 
kinks could interact only very weakly. With the increasing global 
control parameter, the number of either bands or disks decreased, 
and some bands disintegrated into a number of disks, apparently, 
due to zigzag instability. 

One should recall that the global control parameter $\beta_1$ or 
$\beta_2$, as defined above, is proportional to the number of 
``solitary'' objects $N$. Therefore decreasing $N$ brings the system 
back into the parametric region where the remaining solitary 
structures can be sustained. If the values of $\mu$ and $\nu$ 
correspond to the parametric region where both bands and disks are 
stable without global control, bands typically disappear first as the 
global control becomes stronger. Subsequently, the sustainable 
number of disks decreases, and, eventually, inhomogeneous solutions 
disappear altogether, in agreement with the observations by Middya 
and Luss.

\acknowledgements
This work has been supported by the Fund for Promotion of Research 
at the Technion. I thank Dan Luss for stimulating discussions, and for 
acquainting me with Ref. \cite{luss93} prior to publication.

\newpage
\newpage

\end{document}